\documentclass[preprintnumbers]{revtex4}
\usepackage{graphicx}
\def\CP{$ C \! P$ } 

\begin{document}
\preprint{Snowmass E2-41,Bergen ISSN 0803-2696/2001-05}

% You should use BibTeX and revtex.bst for references
\bibliographystyle{revtex}

% number  and your conference paper identification number on the
% title page in preprint mode. Multiple \preprint commands are allowed.
%\preprint{}
\title{Performance of Multipurpose Detectors in Super $B$ Factories}
% Optional argument for running titles on pages
%\title[]{}



\author{G. Eigen}
\email[]{eigen@asfys2.fi.uib.no}
%\thanks{}
\affiliation{University of Bergen}


\date{\today}

\begin{abstract}
Based on background measurements at PEP II 
the impact of machine-related backgrounds on individual 
components of multipurpose detectors is examined, that
operate in an asymmetric $B$ factory
at luminosities up to $10^{36} \ \rm cm^{-2} s^{-1}$. Extrapolations of the
BABAR experience suggests two feasible detector designs. 

\end{abstract}
% \pacs{}

%\maketitle must follow title, authors, abstract and \pacs
\maketitle


% figures should be put into the text as floats.
% Use the graphicx package (distributed with LaTeX2e).
% See the LaTeX Graphics Companion by Michel Goosens, Sebastian Rahtz,
% \begin{figure}
% \caption{}
% \label{}
% \end{figure}

% tables follow here or maybe be put in the text
% \begin{table}
% \caption{}
% \label{}
% \end{tabular}
% \end{table}

%\begin{acknowledgments}
%\end{acknowledgments}

%\bibliography{your bib file}

% You should use BibTeX and revtex.bst for references
\bibliographystyle{revtex}

% number  and your conference paper identification number on the
% title page in preprint mode. Multiple \preprint commands are allowed.
%\preprint{}
\vskip -0.5cm
\section{Introduction}

\CP violation in the $B$ system has been observed recently by 
BABAR~\cite{bbr} and BELLE~\cite{bel}. In a $\rm 29.7~fb^{-1}$ sample
BABAR measured $\sin 2 \beta = 0.59\pm 0.14 \pm 0.05$. In order to achieve
high-precision measurements of all angles in the 
Unitarity Triangle~\cite{burchat}, and to precisely measure 
branching fractions, \CP asymmetries and forward-backward
asymmetries in rare $B$ decays \cite{burchat}, \cite{ge1}
very high luminosities ($\rm 10~ab^{-1}/y$) are a prerequisite.
Recently, an $e^+ e^-$ asymmetric super $B$ factory has been proposed 
that operates at peak luminosities of 
${\cal L}_{peak}=10^{36} \rm \ cm^{-2} s^{-1}$ \cite{seeman}. This is a 
factor of $\sim 250$ higher than presently achieved in PEP II and KEKB.
To maintain $10^{36} \rm \ cm^{-2} s^{-1}$
continuous injection becomes necessary since beam life times are rather short.
This imposes new running conditions on the detector because of 
enhanced machine-related backgrounds.
In the following we examine the performance of 
individual subsystems of present multipurpose detectors 
in a high-radiation environment. Using parameterizations 
that were established in a PEP~II luminosity-upgrade study and are based
on measurements at $3 \times 10^{33}\rm \ cm^{-2} s^{-1}$
\cite{hltf} we extrapolate quantities affected by machine-related backgrounds 
to high luminosities. In addition to results presented in \cite{hltf}
we perform extrapolations to three high luminosity points:
$5 \times 10^{34}\rm \ cm^{-2} s^{-1}$,
$1 \times 10^{35}\rm \ cm^{-2} s^{-1}$ and 
$1 \times 10^{36}\rm \ cm^{-2} s^{-1}$. 
Our results have to be taken with a grain of salt, since extrapolations are 
carried out over 1-2 orders of magnitude and are very specific to the present 
layout of the PEP II interaction region (IR). 
Nevertheless, they are rather instructive and
serve as a crude guideline. 

\section{Background Issues}

Acceptable levels of radiation-induced backgrounds are determined by  
radiation hardness of the detector, occupancy and trigger rate.
Radiation damage causes inefficiencies and eventually leads to the destruction
of detectors. The total integrated dose, that determines the detector
life time, is accumulated under normal running conditions, during
injection, during machine studies and from beam-loss events. 
Injection losses in PEP II contribute only $25-30\%$ of the integrated dose 
in the SVT horizontal plane. Because of the continuous injection 
in the super $B$ factory, it is important to ascertain that the detector
is well-protected from injection losses.
High levels of detector occupancies cause inefficiencies
leading to lower resolutions and reduced signal/background ratios, whereas
high trigger rates cause increased dead times and in turn a loss of signal
events. Thus, acceptable detector occupancies and trigger rates determine
dynamic running conditions of the experiment.

In PEP II  machine-related backgrounds result from
(i) electrons that produce lost particles 
via beam-gas bremsstrahlung and Coulomb scattering in addition to
synchrotron radiation, (ii) positrons that
produce lost-particles via beam-gas bremsstrahlung, Coulomb scattering and
``tune-tails'' and (iii) from
beam-beam interactions, where the latter yields contributions from
luminosity and beam-beam tails in the colliding mode. In the super $B$ factory
beam loss rates are enhanced by a factor of $> 10^3$ as shown in  
Table~\ref{tab:bg} \cite{seeman2}.
Here, the main sources are the Touschek effect and beam-beam tune shifts.
We expect, however, that only a small fraction of the beam losses will 
contribute to backgrounds at the IR. 
For example, in PEP II the LER (HER) beam life time is determined 
by vacuum or the Touschek effect (beam-beam tune shift and second vacuum).
The main background source in the silicon vertex tracker (SVT), the drift 
chamber (DCH) and the electromagnetic calorimeter (EMC), however, comes from
beam-gas interactions in the PEP II incoming straight sections, which have
a negligible effect on beam life times.
Beam losses due to beam-beam tune shifts, dynamic aperture and vacuum
probably will contribute to vacuum-like backgrounds, since they are 
transverse, similarly to distant Coulomb scattering of the LER beam in PEP II.
Since there should be no aperture limitations very close to the IR 
(in particular not from the low-$\beta$ quadrupoles), 
transverse losses are produced at betatron collimators far from the IR. 
Longitudinal losses such as the Touschek effect, whose background impact has 
not yet been studied at PEP~II, probably
behave like distant bremsstrahlung and may be collimated.  
Thus, according to Table~\ref{tab:bg}
the combined transverse losses are the main issue of which 
$15-20\%$ result from the ring-averaged vacuum in the LER. As in PEP II,
the vacuum pressure should be kept at $<10^{-9} $ Torr within a few
10~m of the IR. 
Since the sum of longitudinal losses, all transverse losses and
injection losses is so large, the background issue from vacuum at the IR is
likely to be minor by comparison.

In order to estimate the background contributions due to different
beam-loss effects and determine parameterizations in terms of beam currents
and luminosity in a super $B$ factory several studies are needed. 
Note that our extrapolations are at best
order-of-magnitude estimates, especially at 
$10^{36}~\rm cm^{-2} s^{-1}$, since we presently have no reliable procedure
to include the effects of high beam losses in a super $B$ factory.

\begin{table} [hbtp] \centering
\medskip
\caption{Background composition in PEP II and a super $B$ factory 
\cite{seeman2}}
\label{tab:bg}
\begin{tabular} { | l ||c|c|c|c| }  \hline 
 & HER  & LER & super HER & super LER \\ 
\hline \hline
Beam current $I_b$ [A] & 0.7 & 1.4 & 5.5 & 20.5 \\ \hline
Beam life time $\tau_b$ [min]& 550 & 150 & 4.2 & 3.2 \\ \hline \hline
Luminosity  [A/min] &  & & 0.37 & 0.35 \\ \hline
Vacuum  [A/min] &  & & 0.06 & 0.68 \\ \hline
Touschek  [A/min] &  & & 0.06 & 2.28  \\ \hline
beam-beam tune shift  [A/min] &  & & 0.55 & 2.55  \\ \hline
Dynamic aperture   [A/min] &  & & 0.28 & 1.03  \\ \hline \hline
Total beam loss rate $I_b /\tau_b$  [A/min] &  
$1.3 \times 10^{-3}$ & $9.3 \times  10^{-3}$ & 1.3 & 6.4 \\ \hline
\end{tabular} 
\end{table}

\section{Silicon Vertex Trackers (SVT)}

Silicon vertex detectors, located closest to the beam,  
are exposed to the highest levels of radiation. The specific radiation dose 
depends both on beam currents and on the IR layout. In BABAR the 
horizontal plane is most affected due to the PEP II beam optics at the IR. 
Here the dose rate is a factor of $\sim 5$ larger than in other SVT regions. 
Figure~\ref{fig:vtx} shows the presently accumulated dose rate in the BABAR
SVT. The Si detectors are expected to survive at least 
a 2~Mrad radiation dose. Thus,  
with replacements of detectors in the horizontal 
plane the BABAR SVT should survive luminosities of ${\cal L} = 1.5-3.0
\times 10^{34} \ \rm cm^{-2} s^{-1}$. 
Dose rates in the horizontal plane (forward-east and backward-west) scale with
LER and HER beam currents (in [A]) as \cite{hltf}:

\vskip -0.5cm
\begin{equation}
\begin{array}{rcl}
\rm
D_{SVT} \ [kRad/y]= 128 \cdot I_{LER} + 16 \cdot I^2_{LER} \ \ \ \ \ (for \ 
FE \ MID \ plane), \\
\rm
D_{SVT} \ [kRad/y]= 246 \cdot I_{HER} + 9.1 \cdot I^2_{HER} \ \ \ \ (for \ 
BW \ MID \ plane).
\end{array}
\end{equation}

\vskip -0.1cm
\noindent
The dose rates in the horizontal plane projected for different luminosities are
plotted in Figure~\ref{fig:vtx} and are summarized in Table~\ref{tab:hl}. 
$R \& D$ studies in ATLAS \cite{andricek} have demonstrated that silicon-strip
detectors can survive very high radiation levels in hadronic environments. 
However, this requires cooling and the use of $\rm p^+nn^+$ detectors, which
are read out on the $\rm p^+$ side. The radiation eventually
changes n-type Si into p-type Si moving the depletion layer away from
the readout plane which costs a factor of $\sim 2$ in signal/noise 
\cite{lutz}. This effect also occurs at a reduced rate
in an electromagnetic radiation environment. 
At ${\cal L} = 10^{36} \ \rm cm^{-2} s^{-1}$
occupancy is an issue for Si strip detectors. Thus, the first two layers 
need to be made of pixels. H. Yamamoto succeeded to bond $ 150 \ \mu \rm m$
thick pixels ($\rm 55 \ \mu m\times 55 \ \mu m $) using CMOS technology
\cite{yamamoto}. For the three outer layers Si strip detectors are fine.
To determine the specific layout of the Si pixel and
Si strip detectors, $R \& D$ studies are necessary.
With the appropriate design
Si detectors are expected to work in high-radiation environments.

{\textwidth 10cm
\begin{figure}
\includegraphics[width=6.5 cm]{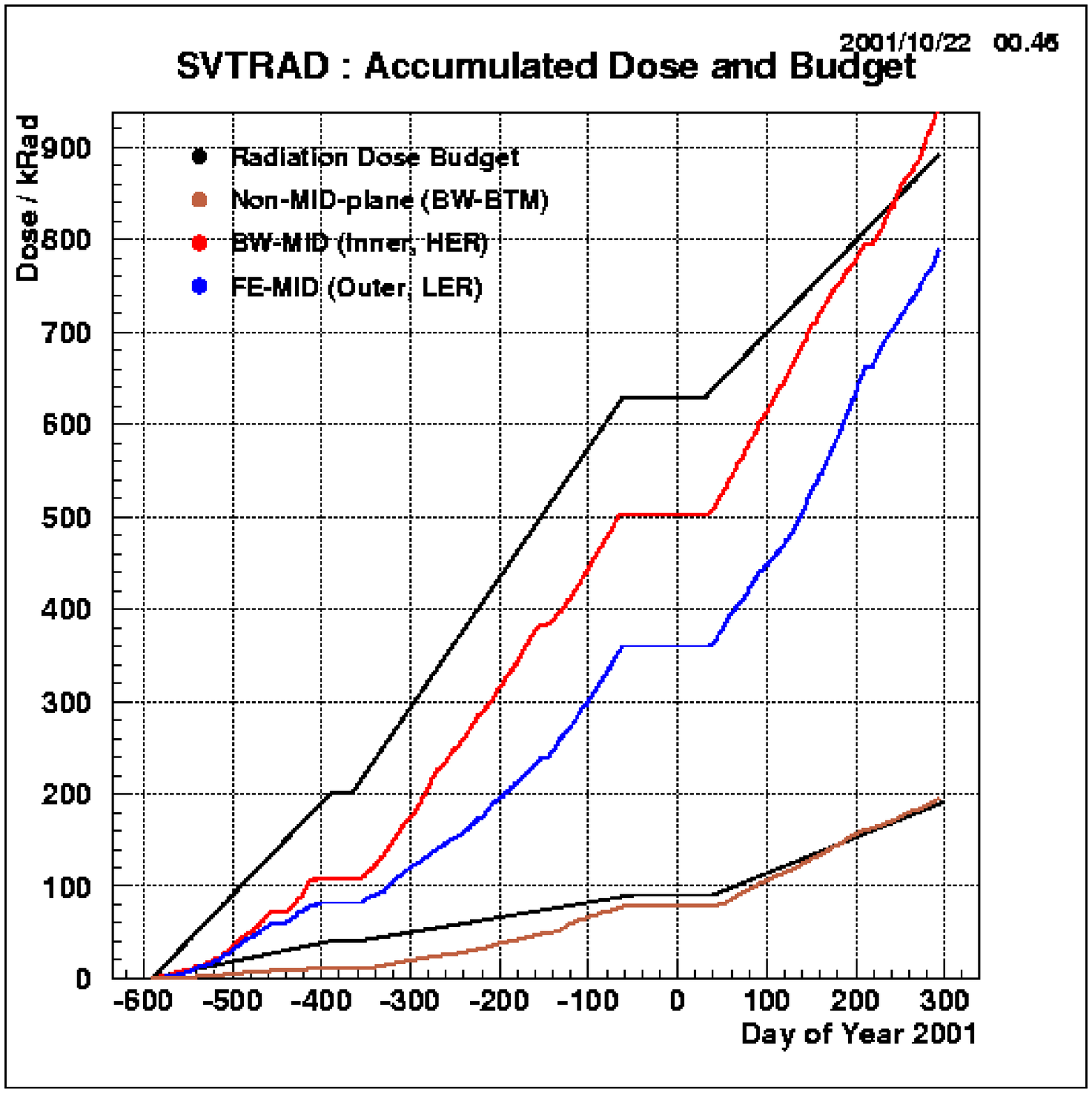}%
\includegraphics[width=7 cm]{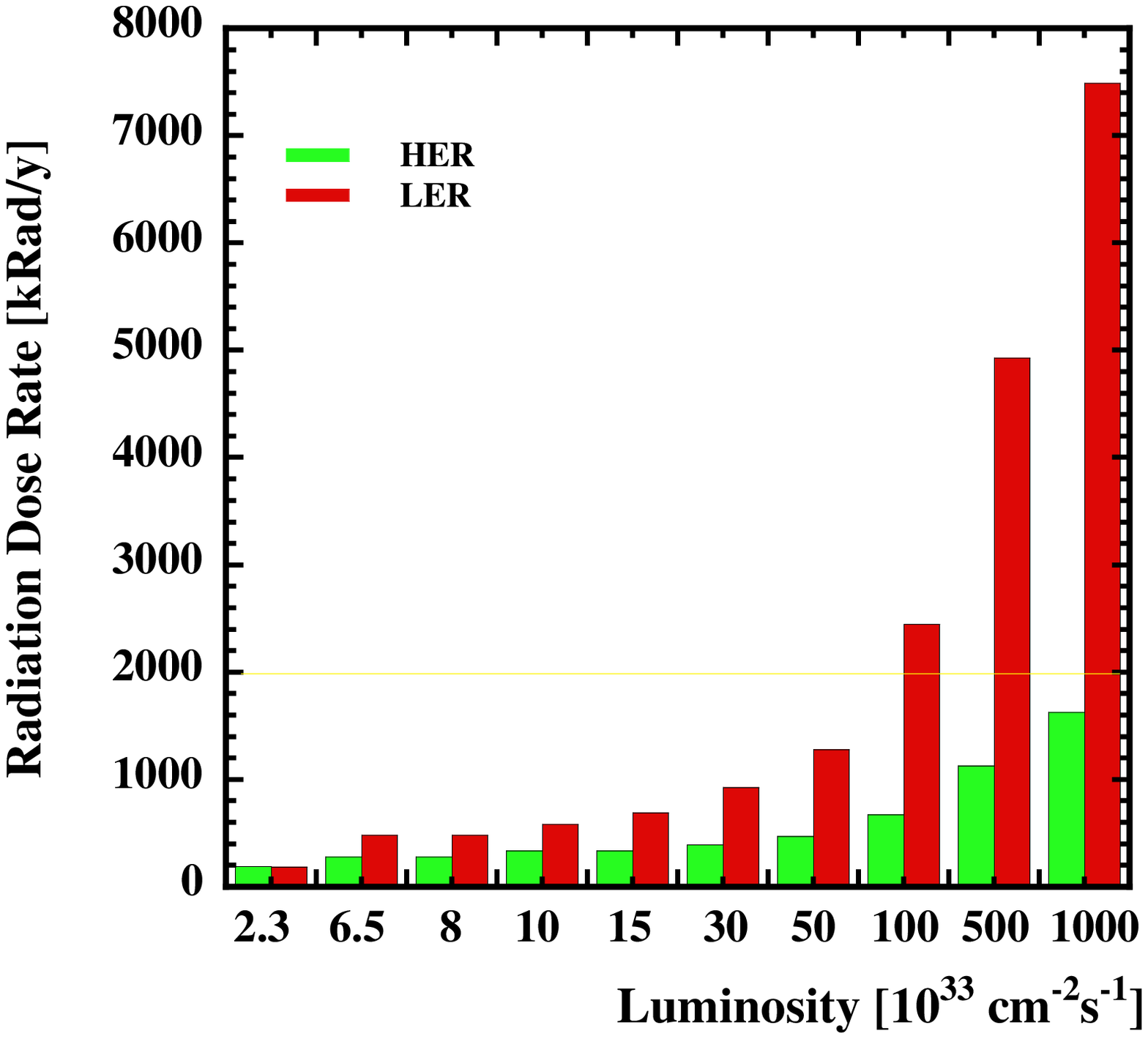}%
\caption{Dose accumulated in BABAR in the horizontal plane and other regions
(left); projected dose rates in the horizontal plane for different ${\cal L}$
(right). The solid horizontal line shows the 2~Mrad dose limit for the BABAR
SVT.
}
\label{fig:vtx}
\end{figure}
}

\section{Drift Chambers (DCH)}

Machine backgrounds affect the operation of a drift chamber in three ways.
First, the total current $I_{DCH}$ drawn by the wires in the drift chamber is
dominated by the charge released by beam-related showers and is limited 
by the high-voltage system. Above this limit
the chamber becomes non-operational. Though this limit can be increased by
adding power supplies, high currents also contribute to drift chamber
aging. Permanent damage is expected to occur at 
charge-densities of $Q_{max} > 0.1\ \rm C/cm$ of wire. Second, 
the occupancy in the drift chamber due to backgrounds can hamper the 
reconstruction of physics events. Third, ionization radiation can permanently 
damage read-out electronics and digitizing electronics. 

Figure~\ref{fig:dcha} shows the DCH currents measured in BABAR as a function 
of LER and HER beam currents for an operating high voltage of 
$U = 1900~\rm V$. The best description of the high-current region is 
achieved with a linear plus quadratic fit for $\rm I_{LER} > 700 ~\rm mA$ 
and $\rm I_{HER} > 150~ \rm mA$. To extrapolate to high luminosities
we use the following
BABAR parameterizations of drift chamber currents and occupancies
in terms of beam currents (in units of [A]) and luminosity (in units
of [$10^{33} \ \rm cm^{-2} s^{-1}$]) \cite{hltf},

\vskip -0.4cm
\begin{equation}
\begin{array}{rcl}
\rm I_{DCH} [\mu \rm A] = 35.3 \cdot I_{LER} + 23.5 \cdot I_{LER}^2 + 
77.2 \cdot  I_{HER} + 46.3 \cdot I_{HER}^2 + 41.9 \cdot {\cal L} -14, \\
\rm N_{DCH} [\%] = 0.044 + 0.191 \cdot I_{LER} + 0.0402\cdot  I_{LER}^2 
+ 1.03 \cdot  I_{HER} + 0.113 \cdot I_{HER}^2 + 0.147 \cdot {\cal L}.
\label{eq:dch}
\end{array}
\end{equation}
\vskip-0.2cm

The drift chamber currents, occupancies and charge densities accumulated on 
the wires extrapolated to high ${\cal L}$
for  $U = 1900~\rm V$ are summarized in Table~\ref{tab:hl}. 
Figure~\ref{fig:dchb} shows the individual contributions of drift chamber 
currents and occupancies due to beam currents and luminosity
extrapolated for different luminosities.
For ${\cal L} > 5 \times 10^{34}\ \rm cm^{-2} s^{-1}$, it
is rather unlikely that a drift chamber will function. Thus, at 
${\cal L} > 1 \times 10^{36}\ \rm cm^{-2} s^{-1}$ other tracking devices
have to be used. Candidates are Si-strip detectors,
straw chambers and GEM detectors. Specific $R \& D$ is needed to find the 
optimal tracker in terms of layouts balancing multiple scattering versus 
position resolution and magnetic field strength versus track length.
For example, a Si strip tracker would be integrated with the SVT. 
Here, multiple scattering is of concern. For a 
combined tracking system with two layers of Si pixel detectors and seven 
layers of Si strip detectors, which are each $200~\rm \mu m$ thick, have
a position resolution of $\rm 20~\mu m$ each and are positioned within a 50~cm 
radius, the momentum resolution in a  3~T magnetic field is 
$\sigma_{p_t}/p_t = 0.47\% \oplus 0.072 p_t \%$ (including a beam pipe). 
The multiple-scattering term is slightly worse 
than that in BABAR \cite{bbr2}, where a momentum resolution of
$\sigma_{p_t}/p_t = 0.45\% \oplus 0.13 p_t \%$ is measured. Including
a mechanical support structure for the Si detectors the 
multiple-scattering term may increase by $20-30\%$.

{\textwidth 10cm
\begin{figure}
\includegraphics[width=12 cm]{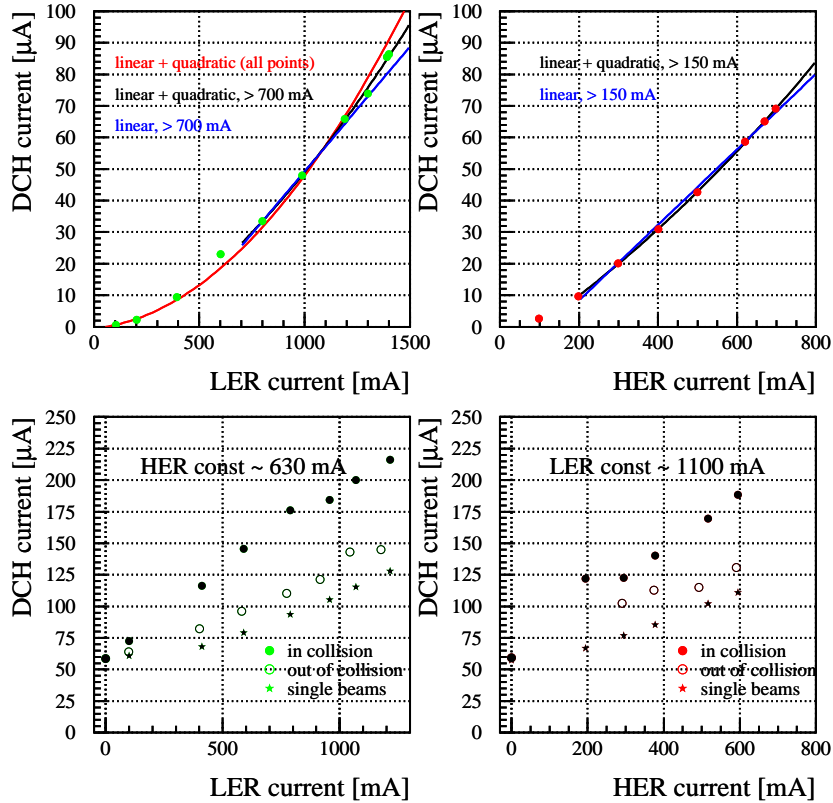}%
\caption{Drift chamber currents as a function of beam currents 
measured in BABAR. The solid lines show different fits.
}
\label{fig:dcha}
\end{figure}
}

{\textwidth 10cm
\begin{figure}
\includegraphics[width=7 cm]{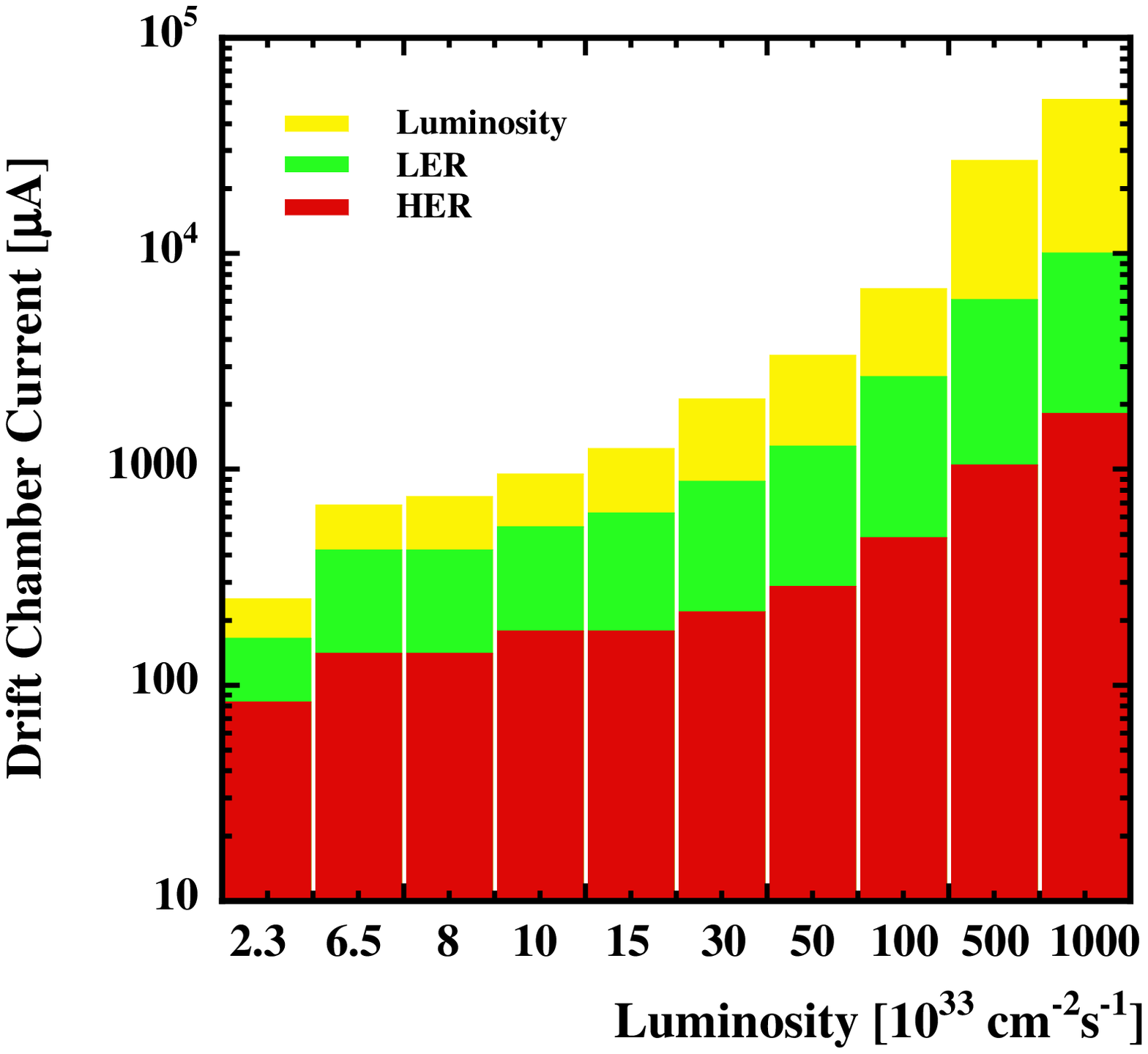}%
\includegraphics[width=7 cm]{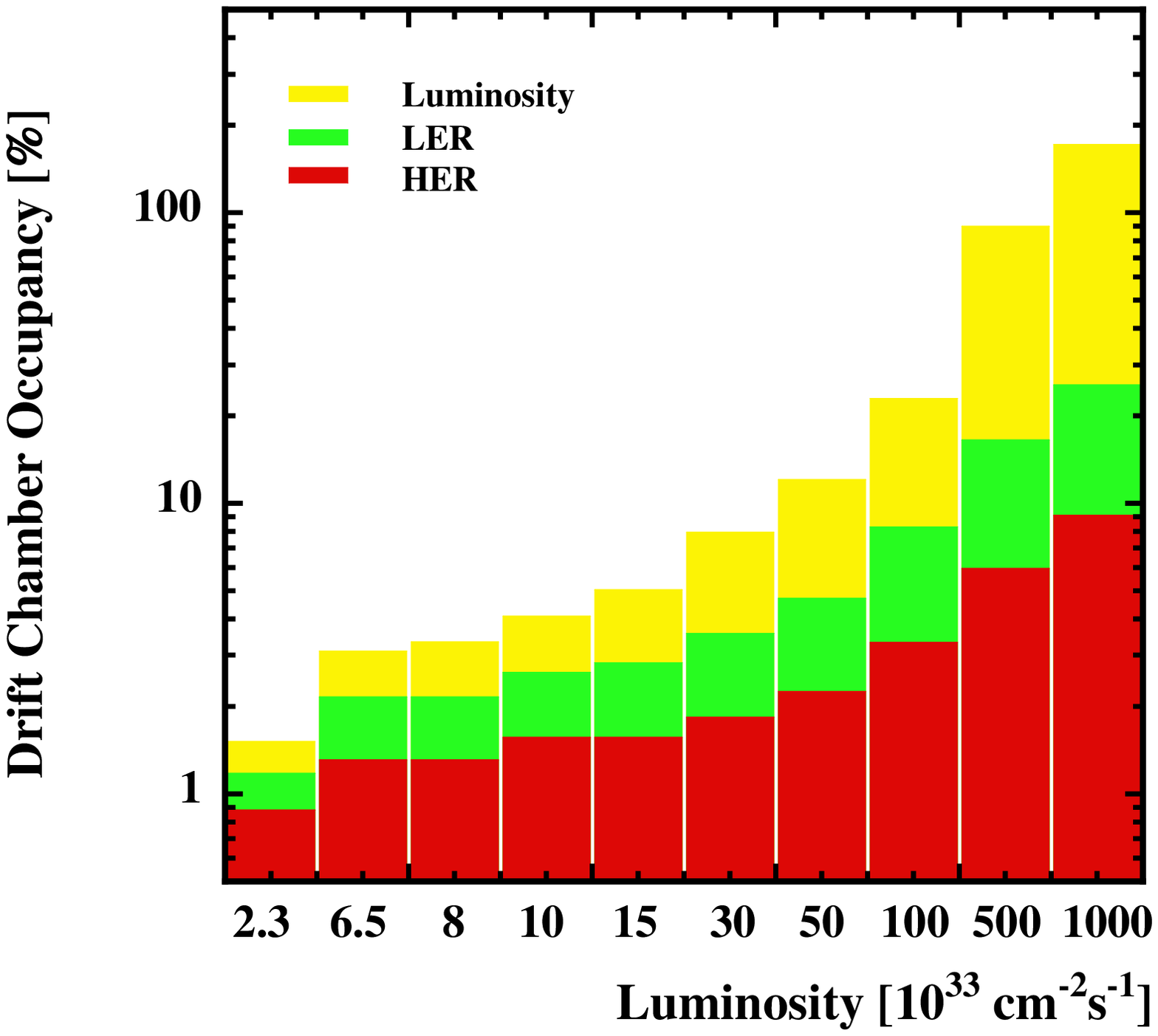}%
\caption{Projected drift chamber currents (left) and occupancy (right) for
different luminosities. Individual contributions due to HER currents, 
LER currents and 
${\cal L}$ are shown in logarithmic scale. Occupancies above 
$100\%$ simply indicate that multiple hits are recorded per wire 
($N_{hits} > N_{wire}$). 
}
\label{fig:dchb}
\end{figure}
}

\section{DIRC Particle Identification}

The main issue for the DIRC is occupancy. In BABAR the occupancy
scales linearly with beam currents ([A]) and luminosity
([$10^{33} \ \rm cm^{-2} s^{-1}$]) \cite{hltf} as,

\vskip -0.5cm
\begin{equation}
\rm N_{DIRC} [kHz] = 35 \cdot I_{LER} + 8.5\cdot  I_{HER} + 25 \cdot {\cal L}.
\end{equation}
\vskip -0.2cm
\noindent
The occupancies extrapolated to high luminosities, assuming detector geometry
and IR layout identical to that in BABAR, are given in Table~\ref{tab:hl}.
Figure~\ref{fig:dirc} shows the DIRC occupancies extrapolated 
for different luminosities indicating the
individual contributions due to beam currents and luminosity.
The DIRC occupancies in BABAR are acceptable up to
${\cal L} < 6 \times 10^{34} \ \rm cm^{-2} s^{-1}$. However, the water
tank provides a huge Cherenkov detector. Thus, at luminosities
${\cal L} > 1 \times  10^{35} \ \rm cm^{-2} s^{-1}$ 
the water tank has to be replaced with a compact readout system based
on focusing and timing. Using pixelated photodetectors with 
a time resolution of $<200$~ps should yield a factor of three improved 
Cherenkov-angle resolutions (2.7~mr) than achieved presently in
the DIRC \cite{burchat}.

{\textwidth 10cm
\begin{figure}
\includegraphics[width=7 cm]{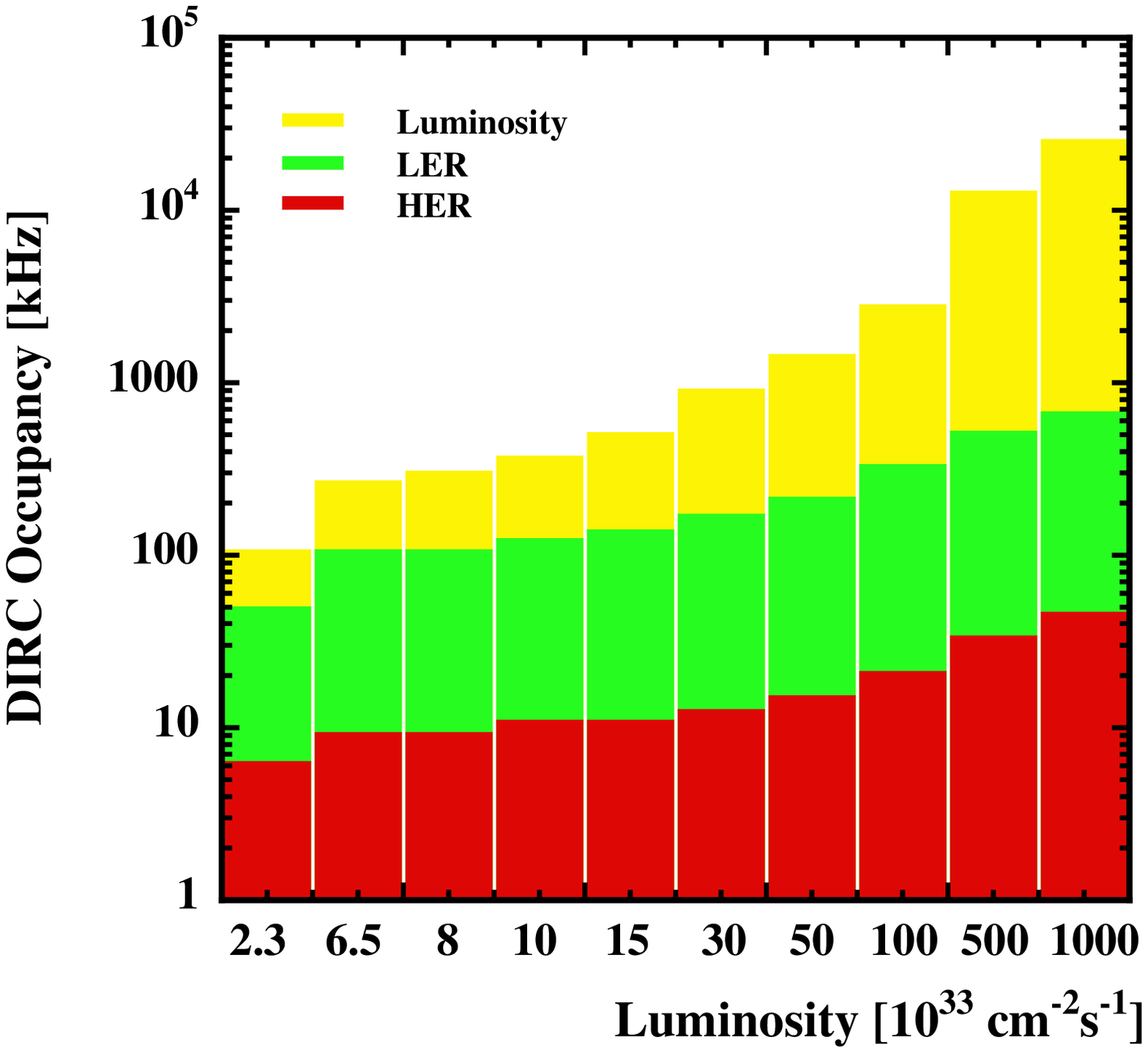}%
\includegraphics[width=6.5 cm]{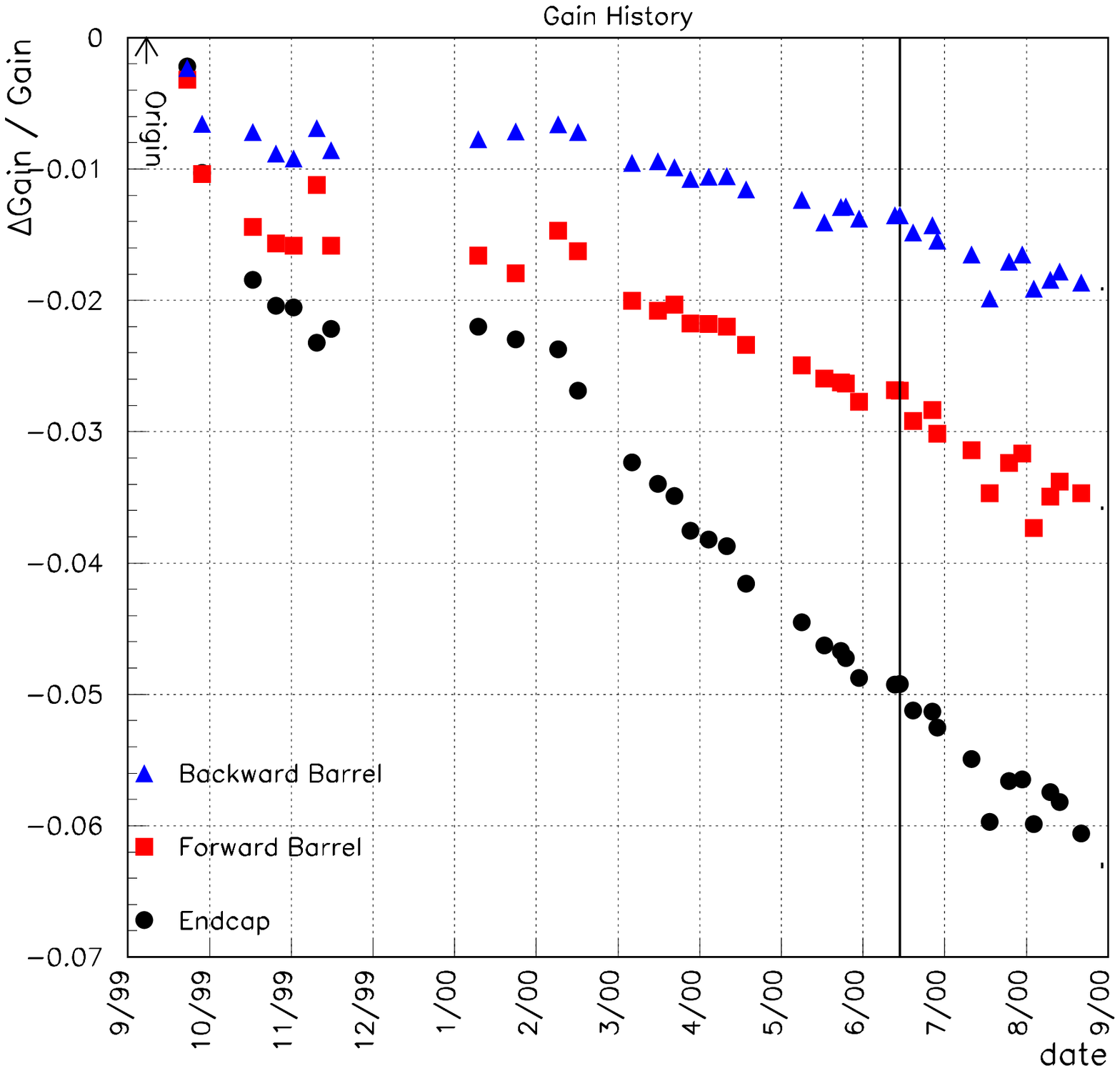}%
\caption{Projected DIRC occupancy for different luminosities showing
individual contributions due to HER currents, LER currents
and ${\cal L}$ in logarithmic scale (left). Observed
light loss in different regions of the BABAR EMC (right).
}
\label{fig:dirc}
\end{figure}
}

\section{Electromagnetic Calorimeter (EMC)}
The performance of a thallium-doped CsI-crystal calorimeter is affected by
radiation damage and occupancy. Radiation damage causes light losses, which
first affect low-energy signals yielding increased inefficiencies. 
In the worst case
the crystals are destroyed. Figure~\ref{fig:emca} shows the 
light-yield changes observed in the BABAR CsI(Tl)-crystal calorimeter. 
The forward endcap is most affected by radiation. At a dose of 160 Rad
accumulated for an integrated luminosity of $\sim 23 \ \rm fb^{-1}$ about
$7.5\%$ of the light is lost. In the forward and backward barrel about a 
100~Rad have been accumulated yielding light losses of $4\%$ and $2\%$,
respectively. 
High occupancies may degrade the energy resolution of real photons 
and increase the combinatorial background in the $\pi^0$ mass spectrum.
Figure~\ref{fig:emca} shows the single-crystal occupancy for 
crystals with a $\geq 1~\rm MeV$ energy deposit and the number of crystals
with energy $\rm E \geq 10~MeV$ in the BABAR EMC as a function of beam
currents. These measurements yield the following parameterizations
in terms of beam currents ([A]) and ${\cal L}$
([$10^{33} \rm cm^{-2} s^{-1}$]) \cite{hltf}:

\begin{equation}
\begin{array}{rcl}
\rm O_{E>1~MeV} [\%] = 9.8 + 2.2 \cdot I_{LER} + 2.2 \cdot I_{HER} + 1.4 \cdot 
{\cal L} \ \ \ \ \ \ \ \ \ \ \ \ \ \ \ \ \ \ \ \ \ \ \   
( > 1~MeV),\ \ 
\\
\rm N_{E>10~MeV} = 4.7 \cdot I_{HER} + 0.23 \cdot I_{HER}^2 + 2.4 \cdot I_{LER}  
+ 0.33 \cdot I_{LER}^2 + 0.6 \cdot {\cal L} \ \ \ (>10~MeV).
\end{array}
\end{equation}

Table~\ref{tab:hl} lists the extrapolated occupancies for different 
luminosities and 
Figure~\ref{fig:emcb} shows the individual contributions due to beam
currents and luminosity. The rapid rise of the
luminosity term underscores the importance of properly handling the
radiative Bhabha debris. 
For ${\cal L} < 1.5 \times 10^{34} \ \rm cm^{-2} s^{-1}$ the integrated 
radiation dose in the BABAR CsI(Tl) calorimeter is unproblematic, 
if the observed light losses scale as expected. 
The impact of the large number of low-energy photons on the EMC 
energy resolution needs to be studied, as $\sigma_E/E$ 
depends on the clustering algorithm, on digital filtering and a low-energy
cut-off. 
Background rates can be reduced by improvements of the vacuum near the 
IR combined with effective collimation against $e^+$ from distant Coulomb 
scattering. For luminosities 
${\cal L} > 1 \times 10^{35} \ \rm cm^{-2} s^{-1}$ light losses due to
radiation damage and occupancy levels in CsI(Tl) crystals are not acceptable.
$R \& D$ is needed to find another inorganic scintillator that works
in this environment. Candidates include
pure CsI crystals read out with APS's, LSO or GSO read out with photodiodes
or APD's. For example, LSO has interesting properties: 
a light yield of 27000 photons/MeV (compared to 56000 photons/MeV in CsI(Tl)),
a radiation length $X_0= 1.14$~cm, a Moli\`ere radius of 2.3~cm, and
radiation hardness of $100$ MRad \cite{burchat}. Though LSO is a factor $>2$
more expensive than CsI(Tl), LSO crystals may be cheaper than CsI(Tl)
crystals, since the volume for the same $X_0$ and same angular segmentation
is a factor of $\sim 3.5$ smaller.

{\textwidth 10cm
\begin{figure}
\includegraphics[width=9 cm]{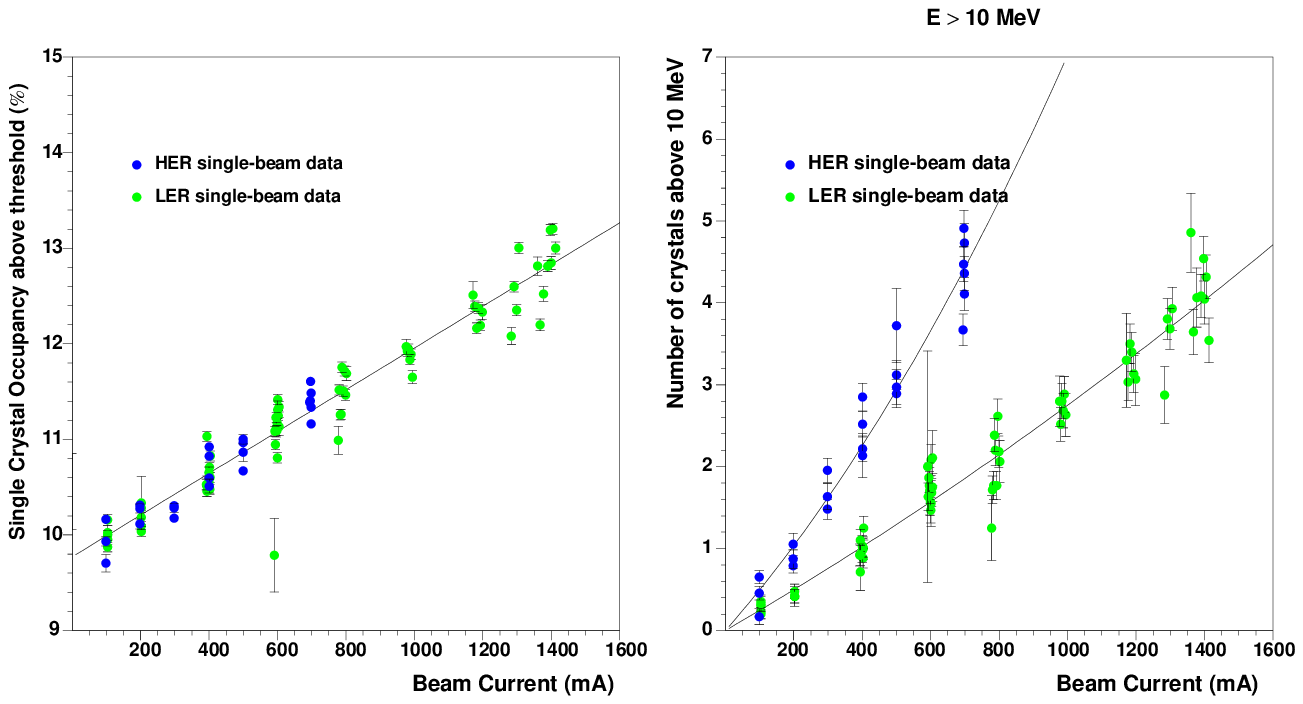}%
\includegraphics[width=9 cm]{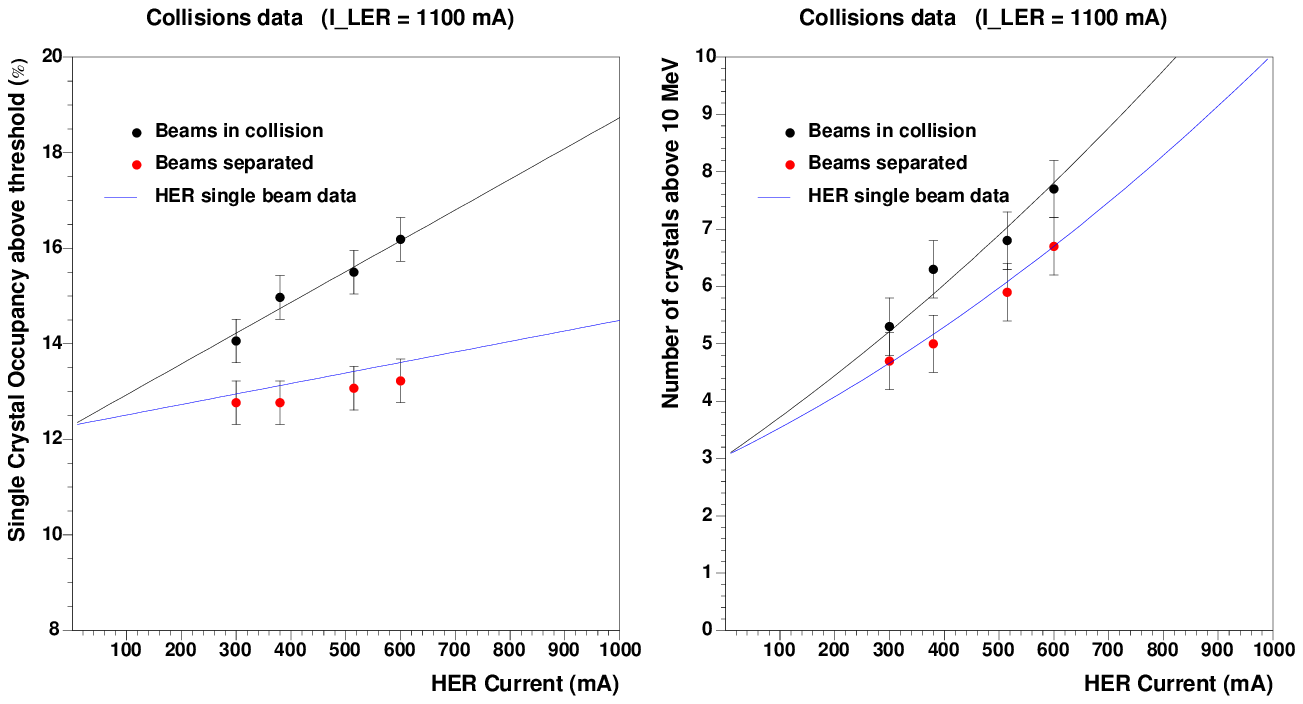}%
\caption{Single crystal occupancy for energy deposits of $\rm > 1~MeV$ and 
number of crystals with energies above 10~MeV observed in BABAR
as a function of PEP II beam currents for single beams (left set) and
two beams (right set). 
}
\label{fig:emca}
\end{figure}
}

{\textwidth 10cm
\begin{figure}
\includegraphics[width=7 cm]{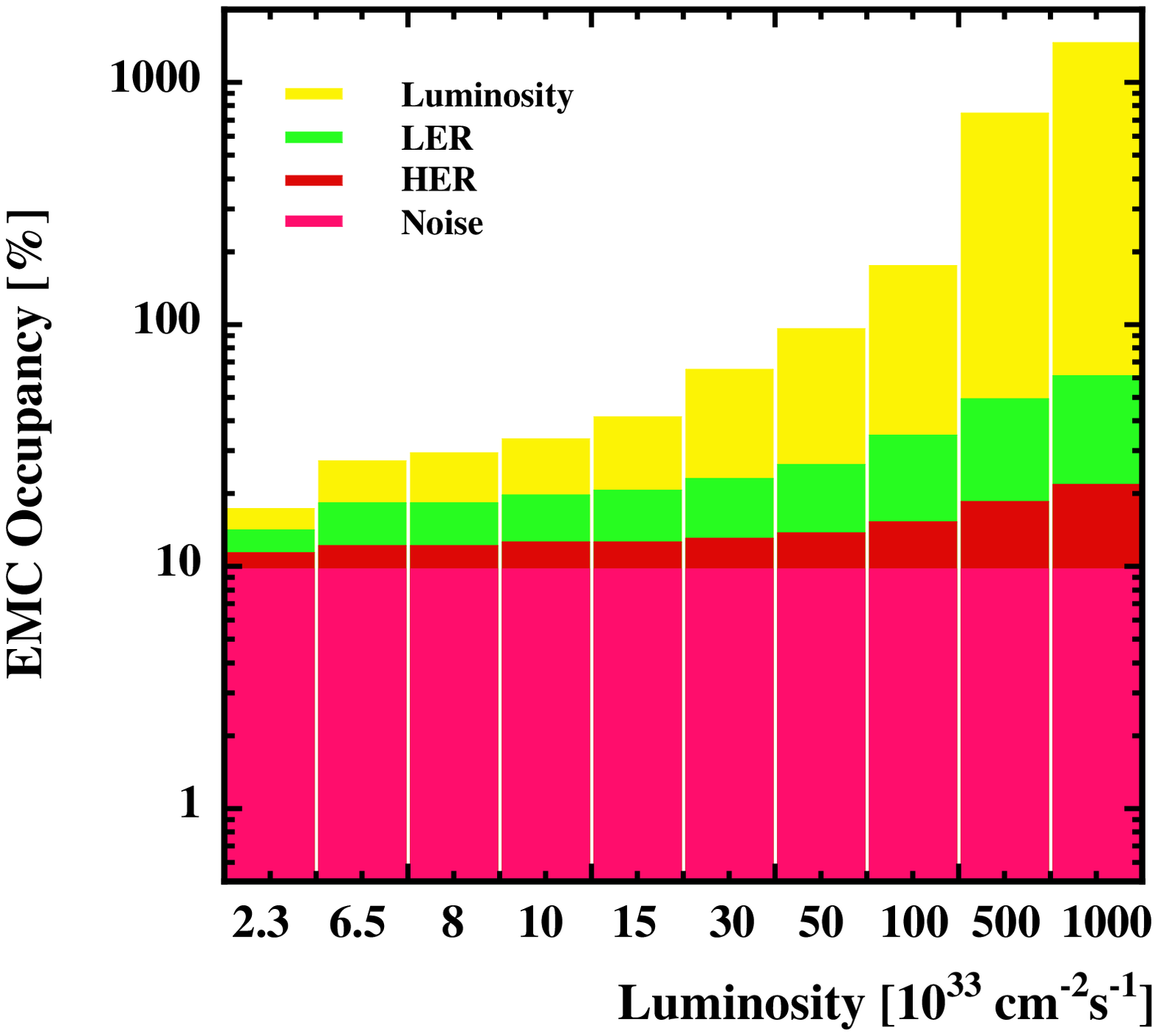}%
\includegraphics[width=7 cm]{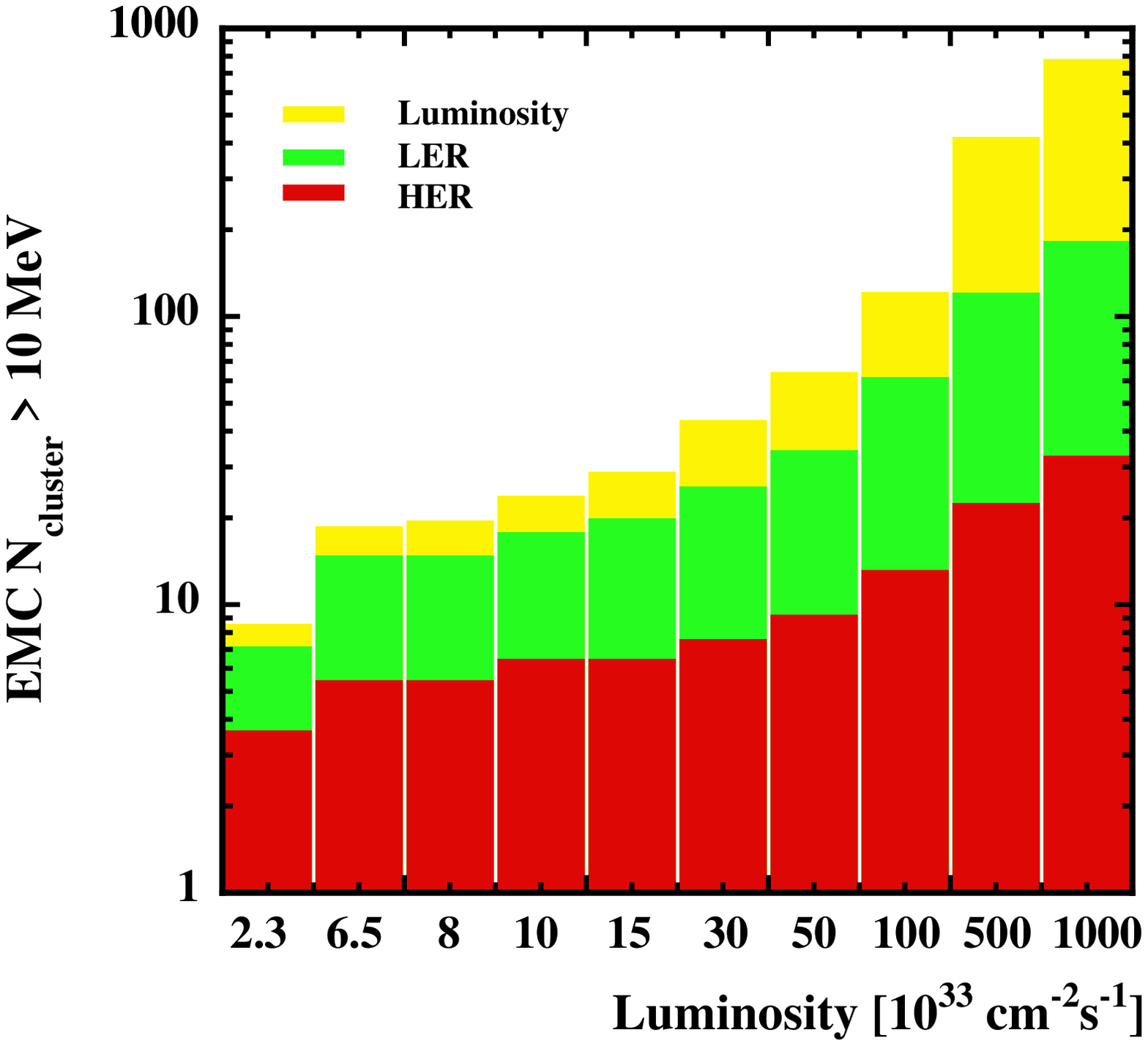}%
\caption{Single crystal occupancy for $\rm > 1~MeV$ energy deposits 
(left) and number of crystals with energies
$>10$~MeV (right) projected for different luminosities. Individual 
contributions due to noise, HER current, 
LER current and ${\cal L}$ are shown in logarithmic
scale. Occupancies above $100\%$ indicate multiple $> 1$~MeV showers 
overlapping in crystals ($N_{shower} > N_{crystal}$).
}
\label{fig:emcb}
\end{figure}
}

\section{Instrumented Flux Return (IFR)}

The main issue for the IFR is the high occupancy in outer layers due to
beam-related backgrounds. In BABAR, for 
${\cal L}=3 \times 10^{33} \ \rm cm^{-2} 
s^{-1}$ the outer RPC layer has an occupancy of several $\%$.
At PEP II design currents this will already become unacceptable due to
an enhanced $\mu/\pi$ misidentification. The solution is to build a 5 cm
thick Fe shield behind the outermost layer of RPC chambers. This 
yields acceptable background rates for luminosities up to 
${\cal L}= 3-5 \times 10^{34} \ \rm cm^{-2} s^{-1}$. Above 
${\cal L}= 1 \times 10^{35} \ \rm cm^{-2} s^{-1}$ occupancy levels become
an issue despite the shielding. Scintillating fibers provide an alternative
readout that may be considered here.

\section{First-Level Trigger Rates (L1)}

The first-level (L1) trigger rates in BABAR scale linearly with beam 
currents and luminosity as shown in Figure~\ref{fig:trig}.
The data can be parameterized by \cite{hltf}

\begin{equation}
\rm L1 [Hz] = 130 \ (cosmics) + 130\cdot I_{LER} + 360 \cdot I_{HER} +  
70 \cdot {\cal L},
\end{equation}

\noindent
where beam currents and luminosity are given in units [A] and
[$10^{33} \ \rm cm^{-2} s^{-1}$], respectively. 
For high luminosities the last term dominates. 
The rates expected at high 
luminosities are listed in Table~\ref{tab:hl}. The different contributions
to L1 are plotted in Figure~\ref{fig:trig}. BABAR is laid out to accept 
an L1-trigger rate of 2.0-2.5~kHZ, depending on the event size. 
To operate BABAR at 
${\cal L} > 1.5  \times 10^{34} \ \rm cm^{-2} s^{-1}$ the L1 trigger needs 
to be upgraded. For luminosities 
${\cal L}> 1 \times 10^{35} \ \rm cm^{-2} s^{-1}$ a new trigger design is 
needed in order to accept the entire physics rate from $b \bar b $ and
$c \bar c$ processes. This is important for analyses that require the full
reconstruction of one $B$ meson. However, we could apply a rather stringent
prescaling of Bhabhas and radiative Bhabhas and reduce beam-gas events by
maintaining a low pressure at and near the IR. At 
${\cal L}= 1 \times 10^{36} \ \rm cm^{-2} s^{-1}$ an L1-tigger rate of 75~kHz
is expected. It should, however, be no problem of building a trigger system 
that can cope with such rates. For example the ATLAS trigger system is 
expected to accept an L1-trigger rate of 100~kHz. This is needed to minimize 
dead times for a 40~Mhz beam crossing rate.

{\textwidth 10cm
\begin{figure}
\includegraphics[width=5 cm]{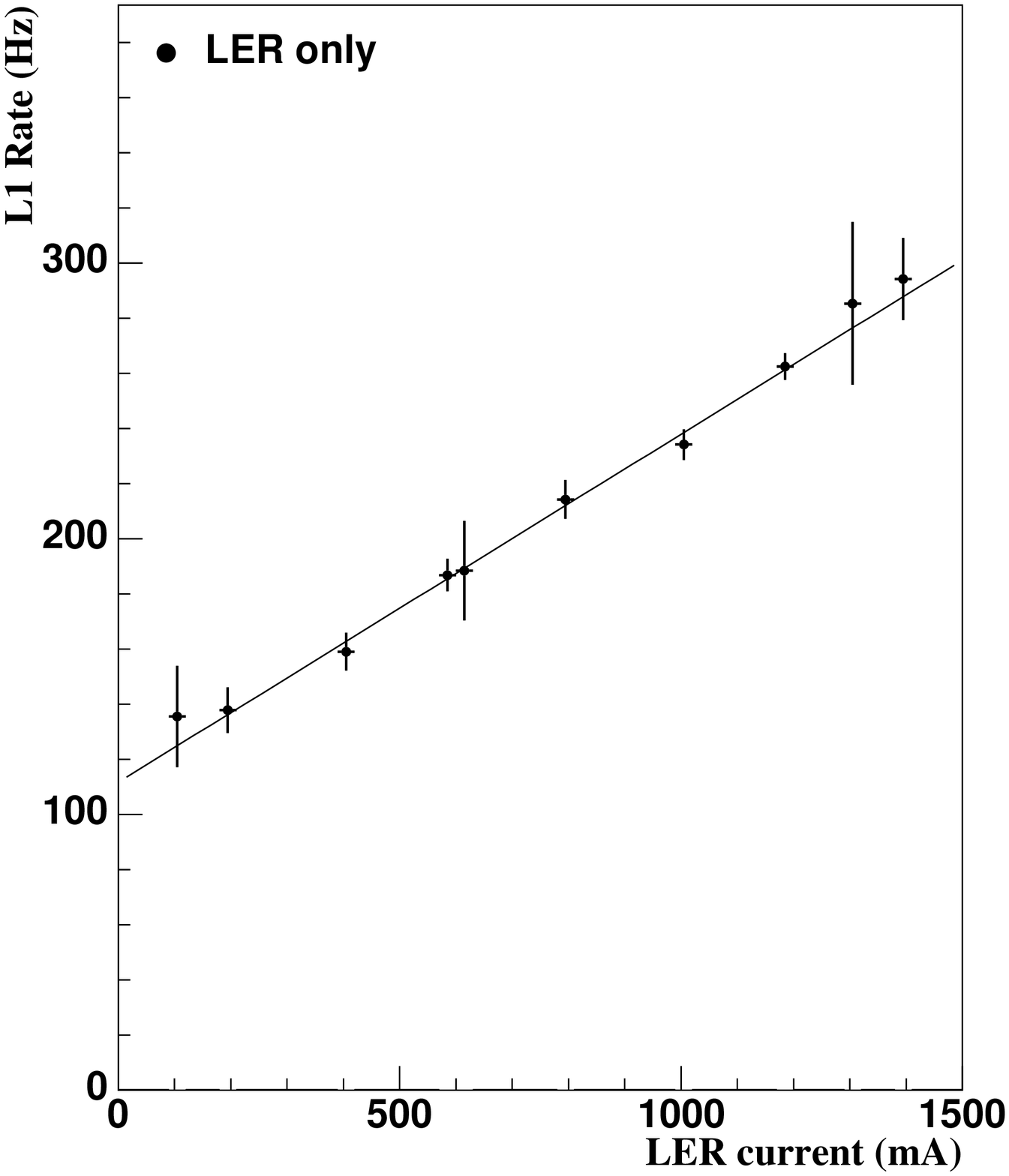}%
\includegraphics[width=5 cm]{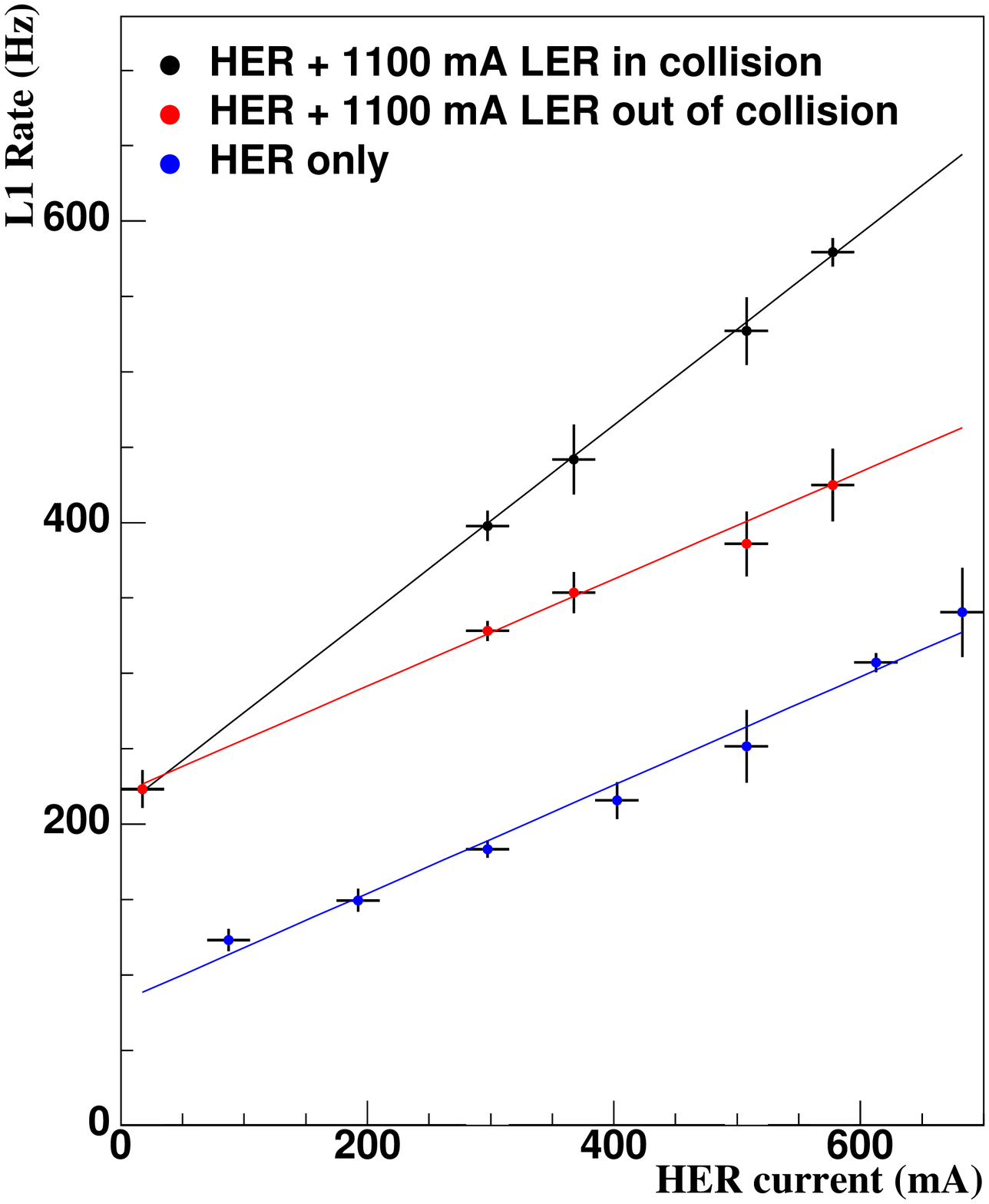}%
\includegraphics[width=7 cm]{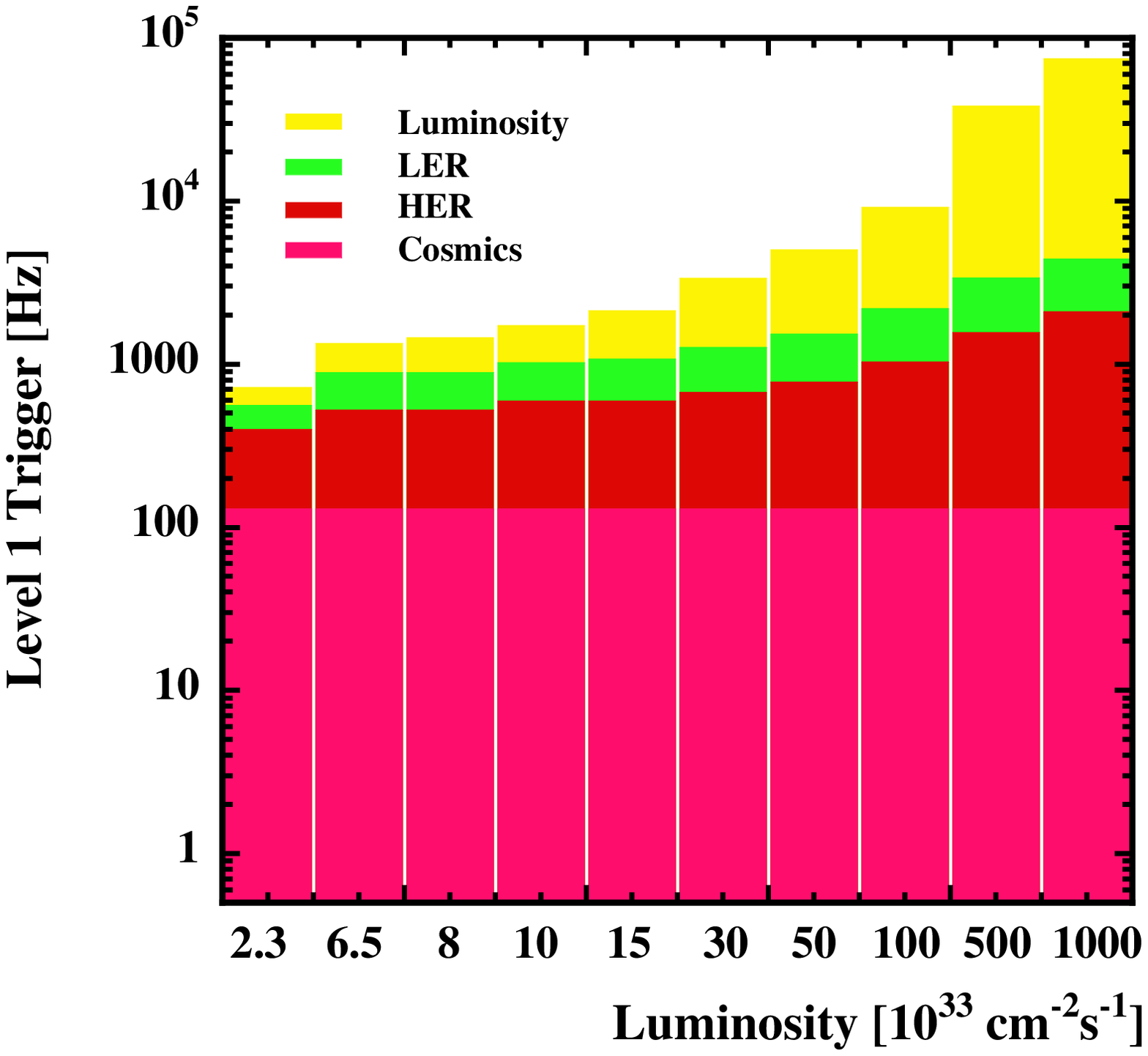}%
\caption{L1 trigger rate in BABAR versus LER (left) and HER
beam currents (center); L1 trigger rates expected for different luminosities 
(right), showing individual contributions due to cosmic rays,
beam currents and ${\cal L}$
in logarithmic scale. 
}
\label{fig:trig}
\end{figure}
}

\vskip -0.3cm
\section{Detector Considerations}

Based on our present knowledge we have considered two detector scenarios:
an upgraded BABAR detector and a new compact detector. In the first 
scenario we would keep the BABAR magnet, the flux return and the DIRC. 
The flux return would be instrumented with scintillating fibers and the DIRC
with a compact readout system using focusing and timing to collect the 
Cherenkov photons in pixelated photon detectors \cite{burchat}.
The SVT, central tracker and electromagnetic calorimeter have to be rebuilt
using one of the options discussed above. Furthermore, new trigger and 
data acquisition systems are necessary. Because of beam-focusing elements
the angular detector acceptance in the super $B$ factory is also limited
to 300~mr as in PEP~II. In order to improve the vertex resolution by a 
factor of two with respect to that in BABAR to achieve an improved separation
of the two $B$ decay vertices for the same boost ($\beta \gamma =0.56$)
as in PEP II, 
the beam pipe radius needs to be reduced to 1~cm. 
For a gold-plated Be beam pipe (1~cm radius, 20~cm length) the expected 
heat load is 0.5~MW from image current heating and 1~MW from incoherent
higher-order modes (HMO) \cite{yamamoto}. Using a double-wall beam pipe with
0.5~mm cooling channels and pumping water at a rate of 1.5 L/min
sufficient cooling is achieved.

\begin{figure}
\includegraphics[width=13.9cm]{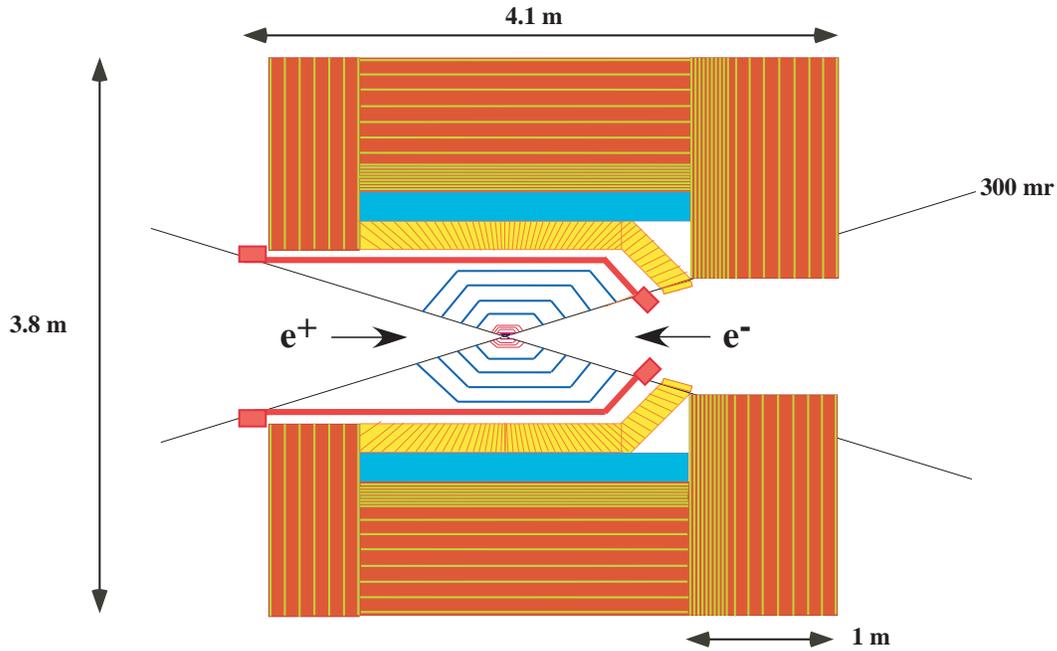}%
\caption{Side view of a compact multipurpose detector for a super $B$ factory. 
}
\label{fig:det}
\end{figure}

A possible choice of a compact detector which also utilizes the asymmetric 
beam energies is sketched in Figure~\ref{fig:det}. 
The tracking systems resides within a 50~cm radius. The SVT consists of
a 2-layer Si-pixel plus a 3-layer Si-strip configuration and is followed
by a 4-layer Si-strip central tracker. The SVT layers lie within 10~cm
starting at a radius of 1.3~cm. All Si layers are bent inwards in the 
forward and backward directions to reduce the path length in the silicon. 
The DIRC starts at 50~cm and consists of a barrel and a forward endcap,
which are respectively read out in the back and the front using a compact
readout system. The EMC starts at $\sim 60$~cm.
To maintain the same position resolution as in BABAR a scintillator with
a reduced Moli\`ere radius is chosen, such as LSO which also has a shorter 
$X_0$ than CsI. The crystals may be segmented into two pieces
to obtain longitudinal shower information which helps to separate hadron 
split-offs from photon showers. The crystals are read out with photodiodes or
with APD's. To compensate for the short track length a magnetic
field of 3 T is considered. The IFR starts at about 100~cm and stretches
to 190~cm. It consists of eight 2.5 cm thick Fe layers interspersed with 
layers of scintillating fibers followed by seven 10~cm thick Fe layers
interspersed with layers of scintillating fibers. In the beam direction the
detector is 4.1~M long.

\begin{table} [hbtp] \centering
\medskip
\caption{Extrapolations of dose rates, currents and occupancies in detectors
for different luminosities.}
\label{tab:hl}
\begin{tabular} { | l || l ||c|c|c|c|c| }  \hline 
Machine &
${\cal L}_{peak} \ [\rm cm^{-2} s^{-1}]$ &$6.5 \times 10^{33} $&
 $1.5 \times 10^{34} $& $5. \times 10^{34} $& $1 \times 10^{35} $&
 $1 \times 10^{36} $ \\ \hline 
properties &
$\int {\cal L} dt \ [\rm fb^{-1}/y]$ &65 & 150 & 500 & 1000& 10000 \\ \hline
&$I_{LER}/I_{HER}$ [A] & 2.8/1.1 & 3.7/1.1 & 4.6/1.5 & 9/2.5 & 18/5.5
\\ \hline \hline
SVT &Dose $D_{SVT} \ [\rm kRad/y]$ for FE-MID/BW-MID
& 480/280 & 690/340 & 1300/470 & 2450/670 &
7490/1630 \\ \hline \hline
DCH &$I_{DCH}\ \rm [\mu A] $& 680 & 1250 & 3370 & 6880 & 51960 \\ \hline
&$N_{DCH} =N_{hits}/N_{wires}\ \rm [\%]$& 3.1 & 5 & 12 & 23 & 173 \\ \hline
&$Q_{wire} \ \rm [m C/cm] $ & $\sim 15$ & 36 & 100 & 200 & 2000 \\ 
\hline \hline
DIRC &$N_{DIRC} \ \rm [kHz] $ & 270 & 516 & 1470 & 2840 & 25700  \\ 
\hline \hline
EMC &$N_{EMC}=N_{shower}(>1~MeV)/N_{crystal}
 \ \rm [\%] $& 28 & 42 & 93 & 175 & 1460 \\ \hline 
&$N_{cluster}\  $ & 21 & 32 & 56 & 122 & 783  \\ \hline \hline
Trigger &$L1 \rm\  [Hz]$& 1350 & 2130 & 4800 & 9200 & 74500 \\ \hline \hline
\end{tabular} 
\end{table}

\vskip -0.3cm
\section{Conclusion}

A multipurpose detector can be built with existing or new technologies
which exploits the unique physics opportunities at a super $B$ factory 
operating at a luminosity  of ${\cal L} = \rm 1 \times 10^{36}~cm^{-2} s^{-1}$.
For example, by performing precision measurements of rare decays \cite{ge1}
the super $B$ 
factory is complementary to searches for New Physics in high-energy colliders.
The detector requirements include good momentum resolution, high
efficiency for low-momentum tracks, good photon-energy resolution, good
position resolution of secondary vertices,
a high $K/\pi$ separation and a high $\mu$ efficiency with a
low misidentification probability.
To find optimal detector choices and construct a detector in a reasonable
time scale $R \& D$ needs to be conducted in several areas soon.

\vskip -0.8cm
%\subsubsection{}

\begin{acknowledgments}
This work has been supported by NFR. I would also like to thank W. Kozanecki 
for useful discussions.
\end{acknowledgments}


\bibliography{e2-41-ge}

\end{document}